# Metallic glasses for spintronics: anomalous temperature dependence and giant enhancement of inverse spin Hall effect

W. Jiao,[1] D. Z. Hou,[1*] C. Chen,[1] H. Wang,[1] Y. Z. Zhang,[1] Y. Tian,[2] Z. Y. Qiu,[3] S. Okamoto,[4] K. Watanabe,[1] A. Hirata,[1,5,6,7] T. Egami,[4,8] E. Saitoh,[1,9,10] M. W. Chen[1,2,11*]

[1] Advanced Institute for Materials Research, Tohoku University, Sendai 980-8577, Japan

[2] Department of Materials Science and Engineering, Johns Hopkins University, Baltimore, MD 21218, USA

[3] School of Materials Science and Engineering, Dalian University of Technology, Dalian, China

[4] Materials Science and Technology Division, Oak Ridge National Laboratory, Oak Ridge, Tennessee 37831, USA

[5] Mathematics for Advanced Materials-OIL, AIST-Tohoku University, Sendai 980-8577, Japan

[6] Graduate School of Fundamental Science and Engineering, Waseda University, 3-4-1 Ohkubo, Shinjuku, Tokyo, 169-8555, Japan

[7] Kagami Memorial Research Institute for Materials Science and Technology, Waseda University, 2-8-26 Nishiwaseda, Shinjuku, Tokyo, 169-0051, Japan

[8] Department of Materials Science and Engineering and Department of Physics and Astronomy, University of Tennessee, Knoxville, Tennessee 37996, USA

[9] Department of Applied Physics, The University of Tokyo, Tokyo, Japan

[10] Advanced Science Research Center, Japan Atomic Energy Agency

[11] CREST, Japan Science and Technology Agency, Saitama 332-0012, Japan

*To whom correspondence should be addressed;
Email: dazhi.hou@imr.tohoku.ac.jp; mwchen@jhu.edu

**Spin-charge conversion *via* inverse spin Hall effect (ISHE) is essential for enabling various applications of spintronics. The spin Hall response usually follows a universal scaling relation with longitudinal electric resistivity and has mild temperature dependence because elementary excitations play only a minor role in resistivity and hence ISHE. Here we report that the ISHE of metallic glasses shows nearly two orders of magnitude enhancements with temperature increase from a threshold of 80-100 K to glass transition points. As electric resistivity changes only marginally in the temperature range, the anomalous temperature dependence is in defiance of the prevailing scaling law. Such a giant temperature enhancement can be well described by a two-level thermal excitation model of glasses and disappears after crystallization, suggesting a new mechanism which involves unique thermal excitations of glasses. This finding may pave new ways to achieve high spin-charge conversion efficiency at room and higher temperatures for spintronic devices and to detect structure and dynamics of glasses using spin currents.**

## Main text

Spintronics, in which spins replace charges in electronics, offers a futuristic possibility of the high efficiency data storage, information transport and quantum computing (*1*). In particular, the interconversion between a spin current and a charge current *via* spin Hall effect (SHE) and inverse spin Hall effect (ISHE) is crucial for realization of spintronics in many device applications (*2-5*). In both SHE and ISHE effects, a transverse current is created due to the deflection of spin-up and spin-down electrons in opposite directions by means of spin orbit coupling, which essentially shares the same physics with the anomalous Hall effect in ferromagnets (*5,6*). The transverse response usually has a scaling relation with longitudinal electric resistivity (*5-7*). Therefore, increasing electric resistivity by introducing chemical impurities and structural defects can synchronize with enhanced spin charge conversion efficiency (*8*). In contrast, the temperature effect is generally minimal in the spin-charge conversion due to the weak temperature dependence of electric resistivity (*9-12*) and, in fact, has not been expected as an effective factor to significantly improve spin-charge conversion in nonmagnetic materials (*5,8*).

In this study we report that the ISHE of metallic glasses (MGs), a group of disordered metastable metallic materials, shows giant temperature dependence from an onset temperature of ~80-100 K to the glass transition points while the electric resistance changes only marginally. The anomalous enhancement leads to ~2 orders of magnitude higher spin charge conversion efficiency, compared to their crystalline counterparts, at room and higher temperatures. The exotic temperature dependence of the ISHE suggests a new mechanism in spin-charge conversion, which most likely involves the interaction of spins with unique thermal excitations of the structurally disordered materials.

We employed ISHE to measure the spin-charge conversion efficiency of MGs. The ISHE setup is depicted in Figure 1A in which a 10nm-thick MG film deposited on an yttrium iron garnet (YIG) substrate is utilized as the detector of spin current which is injected from the underlying spin-pumping source. The conduction electrons in MGs are spin-polarized through the transfer of angular momentum from the precession of local spins in YIG driven by microwave. The injected spin current $j_s$ is converted to an electric current $j_c$ by ISHE in the MG layer. The established electric field $E_{ISHE}$ is perpendicular to the flow direction of spin current $j_s$ and spin polarization $\sigma$, that is, $E_{ISHE} // j_s \times \sigma$. Multicomponent $Au_{50.5}Ag_{7.5}Cu_{26.5}Si_{15.5}$ and binary $Pd_{81.9}Si_{18.1}$ MGs were selected as the model systems in this study because crystalline Au, Pd and their alloys have been commonly used as the detectors of spin-current conversion. Especially, Au does not suffer from the interference from magnetic proximity and the Nernst effect (*5*). Moreover, Au- and Pd-based MGs have high chemical stability and excellent glass forming ability (*13-15*). The structure of the as-prepared MG films was characterized by high resolution transmission electron microscopy (HRTEM) (Fig. 1C), selected area electron diffraction (SAED) and X-ray diffraction (XRD) (Fig. S1 in Supplementary Materials). No crystal lattice fringes and diffraction peaks can be seen from both the real-space and reciprocal-space images, demonstrating the amorphous nature of the pristine films. Differential scanning calorimetry (DSC) demonstrates that the metastable MGs experience dynamic and structural transitions from a glass to a super-cooled liquid and a stable crystal at the glass transition temperature ($T_g$) and the crystallization temperature ($T_x$) during heating from lower to higher temperatures (Fig. 1B).

We measured the spin pumping and electric resistance of the MG films at temperatures from 20 K to above $T_x$ on heating and subsequent cooling. Fig. 2A and B show the magnetic field $H$ dependence of the transverse voltage $V_{ISHE}$ of Au based MG at a series of representative temperatures during heating and cooling. For each $V$-$H$ signal, a pair of sharp symmetric peak / valley ($V_{ISHE}$) appears at the magnetic field concomitant with the occurrence of ferromagnetic resonance absorption when magnetic field and microwave frequency fulfill the resonance condition of the magnetic mode (Fig. S2 in Supplementary Materials). Accompanying with the reversal of the magnetic field, the sign of electronic voltage reverses as well. These characters are fully in line with the expectation of ISHE and verify that the measured electric voltage from the MG film results from the ISHE (*3*).

Interestingly, the ISHE of MGs exhibits anomalous temperature dependence in a wide temperature range from an onset temperature of about 80-100K to $T_g$. In particular, the $V_{ISHE}$-temperature relation has a direct correlation with the dynamic and structure phase transitions of the MGs. At the lower temperature range from 20K to 80K, the $V_{ISHE}$ of Au-based MG has a vanishingly small change with temperature, which resembles those of crystalline metals of Au, Pd, Pt and Ta (*9-12*). When the temperature is higher than 86K for the Au-based MG and 101 K for the Pd based MG, $V_{ISHE}$ turns to be strongly temperature-dependent and dramatically increases from ~130nV at 80K to the maximum of 2063 nV at 340 K for the Au-based MG (Fig. 2C) and 58.6 nV at 100K to 12000 nV at 400 K for the Pd based MG (Fig. 2D). In particular, a large portion of the curves shows a nearly linear temperature dependence of $V_{ISHE}$ in the range from the onset temperature to a transition temperature of ~0.7$T_g$. In the supercooled liquid region between $T_g$ and $T_x$, the temperature enhanced $V_{ISHE}$ appears to saturate and the $V_{ISHE}$-temperature relation shows a plateau with a weak temperature dependence. Further increasing temperature, the $V_{ISHE}$ sharply decreases from 2063nV at 340K to only 63nV at 400K for the Au-based MG. The critical temperature of 400K corresponds to the crystallization temperature $T_x$ of the glassy film as marked in the DSC curve (Fig. 1B). Thus, the significant drop in $V_{ISHE}$ is caused by the crystallization of the MG as confirmed by HRTEM and SAED (Fig. 1D). Upon gradually cooling the crystallized MG film from 400 K to 20 K, the $V_{ISHE}$ remains at lower values with the insignificant change from ~65nV to ~30nV, similar to other crystalline materials (*9-12*).

To gain more insights of the anomalous temperature dependence of the ISHE of MGs, we conducted cyclic $V_{ISHE}$ measurements of the as-deposited Au-based MG in the temperature range from 20K to 320K. When the ending temperature is set below 250K which corresponds to the upper limit of the linear portion of the $V_{ISHE}$-temperature curve (Fig. 2C), the temperature dependence of $V_{ISHE}$ is fully reversible during multiple temperature cycles (Fig. 3A), suggesting that the anomalous temperature dependence of $V_{ISHE}$ is associated with intrinsic thermal excitations in the disordered materials. When the highest temperature of the cycling is increased to 320K, entering into the nonlinear $V_{ISHE}$-temperature region, $V_{ISHE}$ lags behind the linear temperature dependence and becomes obviously lower during cooling than that measured during heating. The transition temperature, 250 K, from linear to nonlinear response is about 0.75 $T_g$, which is close to the starting temperature of the Johari-Goldstein (or slow *β*) relaxation (*16*). Therefore, the high temperature nonlinear $V_{ISHE}$-temperature dependence is most likely caused by the irreversible structural relaxation with the reduction in the atomic-level stresses and geometric frustration (*17, 18*). The higher $V_{ISHE}$ of the as-deposited glasses than that of the relaxed one indicates that the temperature dependence of the spin-charge conversion is sensitive to local structure of MGs and could be utilized as an experimental tool to probe the structural evolution and dynamics of the disordered system. Importantly, although the high-temperature structural changes obviously affect the slops of the $V_{ISHE}$-temperature relations, the onset temperature remains nearly unchanged, indicating that it is a characteristic temperature for the interaction between spins and the dynamics of glasses.

Note that the anomalous increase in $V_{ISHE}$ at temperatures above the onset temperature cannot be attributed to the variation in charge carrier density in the MG/YIG structure because the ordinary Hall effect remains nearly constant in the temperature range from 20K to 250K (Fig. S3 in Supplementary Materials). The intrinsic magnetoresistance of the MG films is also irrelevant to the anomalous increase in $V_{ISHE}$ because the magnetic field dependence of electric resistance is vanishingly small in the temperature range (Fig. S4 in Supplementary Materials). In stark contrast to the dramatic increase in $V_{ISHE}$ at temperatures above the onset temperature, the electric resistance $R$ of the Au-based MG film decreases only slightly from 75.28 μΩ at 20 K to 74.06 μΩ at 340K (Fig. 2C). The mean free path of electrons in the MG is estimated to be 4.69 Å at 20 K and 4.77 Å at 340 K (Fig. S4 in Supplementary Materials), which is at the same length scale of short-range structural order of MGs and implies that charge transport is governed by strong elastic scattering by structural disorder. The weak linear temperature dependence of $R$ is in line with the asymptotic behavior of the Debye-Waller factor at high temperatures, suggesting the atomic vibration and thus the fluctuations of interatomic distances affect charge transport faintly (*19*). We attempted to fit the plots of the transverse response $V_{ISHE}$ verse longitudinal electric resistivity $R$ of the MGs. Although the small variation range of the resistance is insufficient to warrant a reliable scaling relation, the $V_{ISHE}$ verse $\rho$ plots show a totally different tendency (Fig. 2E) from the prevalent scaling relations established for crystalline metals and alloys (*9-12*), indicating a new ISHE mechanism of MGs.

At the first glance, one may attribute the abnormal inverse spin-Hall response to the disordered atomic structure of MGs in which the inversion symmetry is absent all over and the Rashba spin-orbit coupling may become locally effective, similar to the symmetry-breaking interfaces and surfaces in crystals. However, the near identical ISHE signal of glassy and crystallized samples at the temperatures below the onset temperature of about 80-100K suggests that the static disordered structure itself is not the origin of the anomalous inverse spin-Hall response. In fact, the large temperature dependence of $V_{ISHE}$ with a well-defined onset temperature indicates that the thermal excitations of the disordered MGs play the key role in the anomalous ISHE phenomenon. We measured the specific heat $C_p$ of the Au-based MG by the heat pulse method. Consistent with the known $C_p$-temperature dependence of MGs (*20*), the experimental curve obeys the superposition of the Debye model and the localized Einstein vibrational mode with the Einstein temperature of about 78.3K. The Boson peak as excess vibrational mode over the Debye model is clearly demonstrated in the form of $C_p/T^3$ (Fig. 3B). Apparently, the Boson peak, the well-known low-energy excitation of glasses, does not have direct correspondence with the anomalous temperature dependence of $V_{ISHE}$ in both temperature and frequency.

Different from long-range periodic crystals, MGs contain a multitude of different atomic configurations corresponding to metastable states at local minima of the potential energy landscape (21). The metastable feature offers metallic glasses rich dynamic modes in a broad spectrum (Fig. 3D) and, particularly, there are unique dynamic excitations that are non-existent in crystals but universal in glassy materials (22). The excitation of MGs can be approximately described by a two-level system in which two metastable states have an asymmetric double-well potential with a barrier energy $V_a$, asymmetry energy $\nabla$, and generalized distance $d$ (*23-24*). When the generalized distance $d$ is sufficient small, the barrier between two metastable states can be passed by 'tunneling' through the asymmetry energy $\nabla$ at low temperatures or by thermal activation overcoming the barrier energy $E_a$ at high temperatures (*25*). The $V_{ISHE}$-temperature relations of Au and Pd based MGs before the occurrence of irreversible structural relaxation can be well fitted by a thermal activation equation: $V_{ISHE} = V_c + V_0 \exp(-E_a/K_B T)$, where $V_c$ and $V_0$ are temperature-independent constant and $k_B$ is the Boltzmann constant, based on a simplified two-level model with the assumption that the asymmetry energy $\nabla$ is close to zero and the anomalous enhancement of $V_{ISHE}$ is proportional to the relaxation rate of MGs (Fig. 3C and Fig. S6 in Supplementary Materials). The fitting parameters are summarized in Table S1 (see Supplementary Materials). Both the onset temperatures and linear relationship of the experimental $V_{ISHE}$-temperature curves can be captured by this basic model. The fitted barrier energies of the relaxation are determined to be 29.4meV for Au-based MG and 37.5meV for Pd based one. The corresponding relaxation rate, $\upsilon=\upsilon_0\exp(-E_a/k_BT)$, where $\upsilon_0$ is the atomic vibration frequency and usually in the order of $10^{12}$~$10^{13}$Hz, is estimated to be $10^{10}$-$10^{12}$Hz, which is in line with the known relaxation time of spins in Au (*26*) and locates in the region of fast *β* relaxation in glasses (*23*) as marked by the arrowhead in the schematic hierarchical relaxation spectrum (*27*) (Fig. 3D). Therefore, the anomalous temperature enhancement of ISHE most likely results from the coupling of spins with the fast *β* relaxation of MGs. This assumption is further supported by the consistency between the experimental $V_{ISHE}$-temperature relations and a generalized phonon skew scattering model which is originally developed by Gorini *et al* (*28*) (Fig. S7 in Supplementary Materials).

In addition to the phonon skew scattering model, the anomalous temperature dependence of $V_{ISHE}$ could also be related to reversible local spin fluctuations caused by the localized Dresselhaus type spin-orbit fields (*29*). It is known that local spin fluctuations in spin glasses can enhance $V_{ISHE}$ (*30*). The fast structural relaxation of non-magnetic Pd and Au MGs may create local spin and orbital fluctuations and gives rise to the temperature-dependent $V_{ISHE}$ of MGs. While the underlying physics of the anomalous ISHE requires future experimental and theoretical investigations, it can be concluded that the local structure excitations unique to glasses are responsible to the giant temperature dependence of $V_{ISHE}$ discovered in this study on the basis of the current experiments and preliminary analysis.

In summary, the results presented here suggest that disordered MGs have a unique advantage as a medium for spin-charge conversion. The anomalous temperature dependence of ISHE, resulting from the thermal excitations, is most likely associated with local structural and spin fluctuations in MGs, unique to the glassy state, and thus cannot be described by the scaling law established in crystalline materials. The strategy of using glassy structure for dynamic spin scattering may open a new pathway to design and create novel materials with high spin-charge conversion efficiency for applications in spintronic devices. The strong interaction between spins and thermal excitations of glasses and supercooled liquids may also be utilized to detect the local structure and dynamics of glassy materials and liquids.

**Acknowledgements**

We thank G. E. W. Bauer, N. Nagaosa, Xiaofeng Jin, T. Kikkawa and T. Hioki for valuable discussions. This work was sponsored by JST-CREST "Phase Interface Science for Highly Efficient Energy Utilization", JST (Japan); the fusion research of World Premier International (WPI) Research Center Initiative for Atoms, Molecules and Materials, MEXT, Japan, Japan Society for the Promotion of Science (JSPS) Grant (Kiban-A 17H01325). M.C. was supported by Whiting School of Engineering, Johns Hopkins University. S.O. and T.E. were supported by the US Department of Energy, Office of Science, Basic Energy Sciences, Materials Science and Engineering Division. W. J. was supported by the JSPS postdoctoral fellowship program (ID P15373). We thank the support of JASRI/SPring-8 under Proposals of 2013A1539 and 2013B1197.

**Data and materials availability:** Data is available *via* request to the corresponding authors.

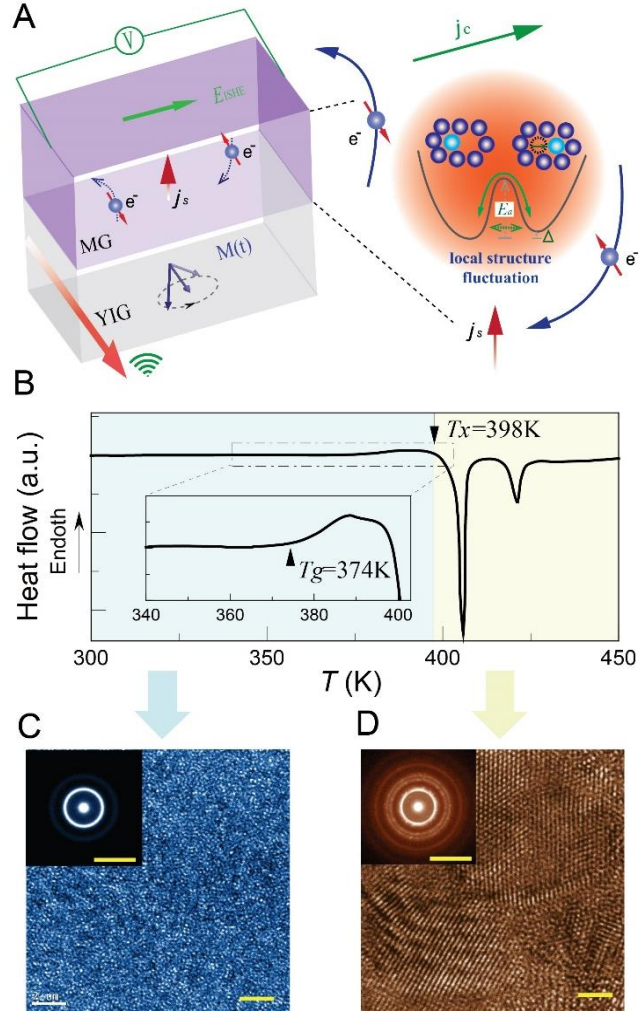

**Figure 1. A**, Schematic illustration of a MG/YIG bilayer system used for the spin-pumping ISHE measurements. Finite temperature excitations in MGs take place through local structural reconfiguration over a rugged potential energy landscape (schematically described as asymmetrical double well potential here) by thermal activation or tunneling. The excitations enhance the spin-charge conversion by dynamic spin orbit coupling. **B,** Differential scanning calorimetry (DSC) curve of the representative Au-based MG with the characteristic glass transition temperature $T_g$ of 374K and crystallization temperature $T_x$ of 398K, as marked by arrows. The DSC measurement was conducted at the heating rate of 5K.min$^{-1}$. High resolution transmission electron microscopy images and the selected area electron diffraction patterns of the pristine Au-based MG (**C**) and the crystallized alloy (**D**).

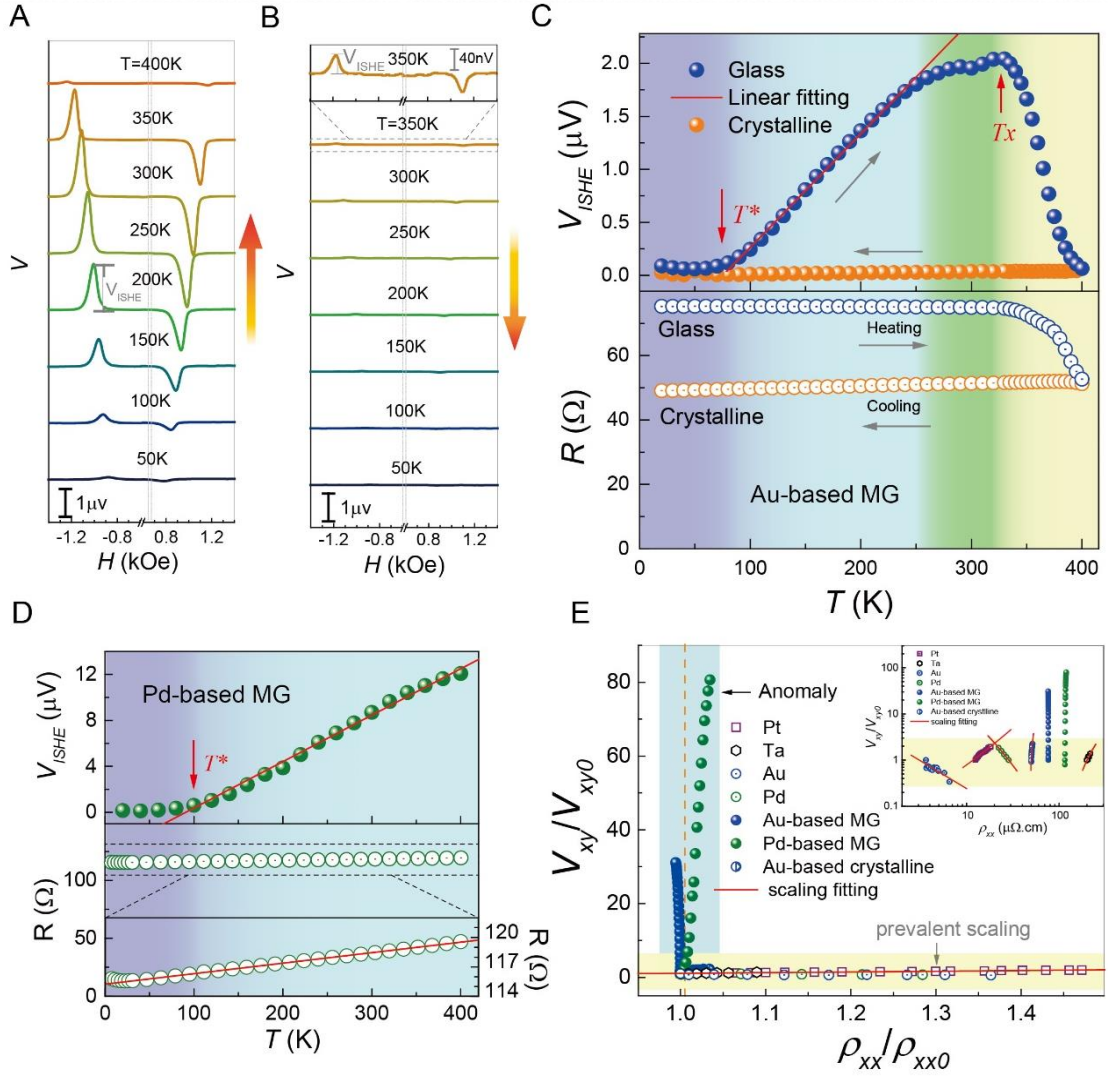

**Figure 2. A,** Magnetic field (*H*) dependence of $V_{ISHE}$ at a series of representative temperatures on heating across the crystallization point (398K) of the Au-based MG. **B**, Magnetic field (*H*) dependence of $V_{ISHE}$ during subsequent cooling of the crystallized MG. The inset at the top is the magnified plot of the 350K curve. **C**, Temperature dependence of $V_{ISHE}$ and *R* of the Au-based MG and the crystallized counterpart. $V_{ISHE}$ of the MG increases dramatically with temperature from about 80K to the crystallization temperature $T_x$, while *R* keeps nearly constant with only 0.61% deduction. **D**, Temperature dependence of $V_{ISHE}$ and *R* for the Pd-based MG. The inset in the middle of the plot is the magnified R-T plot showing the weakly temperature dependence of *R*. **E**, Plots of normalized $V_{ISHE}$ *versus* normalized longitudinal electric resistivity of MGs and crystalline metals and alloys. The dramatic increase of $V_{ISHE}$ in MGs decouples with the minor change of $\rho$. In contrast, the prevalent scaling for crystals: $V_{xy}=a\rho_{xx}+b\rho_{xx}^2$, plotted in the log-log fashion in the inset shows the strong coupling between transverse ISHE response and longitudinal electric resistivity $\rho$.

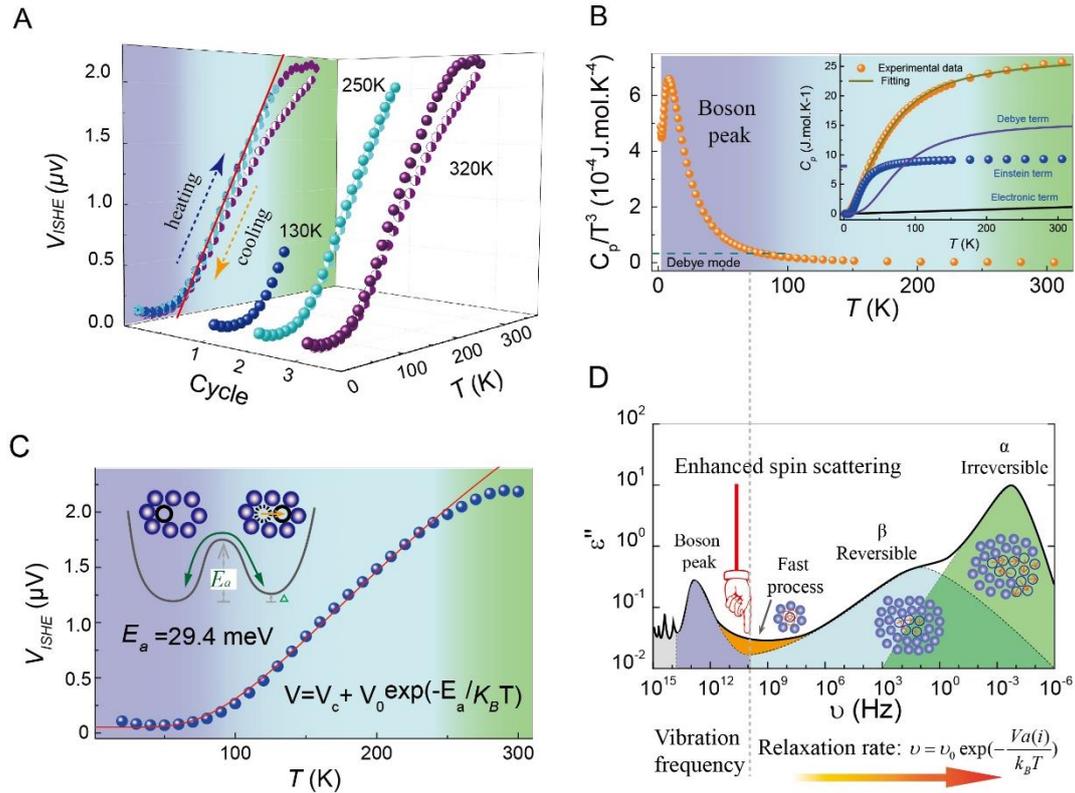

**Figure 3. A**, The temperature dependence of $V_{ISHE}$ during multiple heating-cooling cycles with different ending upper temperatures. The $V_{ISHE}$-T relation is fully reversible when the ending temperatures are lower than $0.75T_g$ (~250K). **B**, Specific heat of Au-based MG. Boson peak as an excess vibrational mode over the Debye model is clearly demonstrated in the form of $C_p/T^3$, and well fitted by the localized Einstein vibrational mode with the Einstein temperature $\theta_E$ of 78.3K shown in the inset. **C**, The fitting of the $V_{ISHE}$-T relation of the Au-based MG by a simplified two-level model. The temperature dependence of $V_{ISHE}$ at the temperatures below $0.7T_g$ is well consistent with the model with the derived barrier energy of 29.4meV. **D**, The associated relaxation rate, $\upsilon=\upsilon_0\exp(-E_a/k_BT)$ ($\upsilon_0$ is atomic vibration frequency usually in the order of $10^{12}$~$10^{13}$Hz), locates in the region of fast β relaxation of glasses as marked by the arrow and dot line in the schematic hierarchical relaxation spectrum.

**Supplementary materials**

Materials and Methods
Table S1
Figures S1 to S7

**Materials and Methods**

**Metallic glass preparation.** ~10 nm thick $Au_{50.5}Ag_{7.5}Cu_{26.5}Si_{15.5}$ (at.%) metallic glass films were sputtered onto the $Y_3Fe_5O_{12}$ substrates at 244K and the working argon pressure of 0.3Pa from a stoichiometric master alloy target. Binary $Pd_{81.9}Si_{18.1}$ (at.%) metallic glass films were deposited with the same sputtering parameters except that the $Y_3Fe_5O_{12}$ substrates were heated to 423K. Cylinder shaped bulk metallic glass samples with a about 2mm diameter were fabricated by copper mould casting. The cylinders were sliced as 1~2mm thick sheets for heat capacity measurement, and about 5mm long rods for measuring thermal conductivity test.

**Structure characterization.** The phase character of the as-deposited glassy films and the final product experiencing *in-situ* temperature varied spin pumping measurement across the crystallization process were characterized via X-ray diffraction (XRD), Cs-corrected high resolution transmission electron microscopy (JEM-2100 F, JEOL) and selected area electron diffraction (SAED). The glassy nature and crystallization behavior were further studied by thermal differential scanning calorimetry (DSC) using Perkin-Elmer Q8500 at a heating rate of 5K/min. The chemical compositions of thick Au- and Pd-based MG films were measured by inductively coupled plasma atomic emission spectroscopy (ICP-AES).

**Physical property measurements.** The electric resistance, Hall effect and magneto-resistance of the 10 nm thick MG films were measured by using Physics Property Measurement system (PPMS, Quantum Design). For specific heat of the MGs, 1~2mm thick MG sheets with nearly identical compositions and structure of 10 nm thick films were used and measured at the temperature range between 2K and 320K using the heat capacity function of the PPMS. Thermal conductivity and electric resistivity were measured simultaneously from 2K to 320K. The MG samples with the dimensions of about 5mm in length and 1mm in diameter were used for the simultaneous measurements of thermal conductivity and electric resistivity at the temperatures from 2K to 320K.

**Spin pumping and electric resistance measurements**. The MG/YIG bilayer systems were placed on the co-planar waveguide of a home-made RF probe compatible with the PPMS. The simultaneous measurements of ISHE voltages and electric resistance were conducted in a wide temperature range from 400 K to 20 K. Five-GHz microwave of 23 dBm is applied to the MG/YIG bilayer systems to generate the spin current at the ferromagnetic resonance fields (H_FMR) and the resulting inverse spin Hall response in the MG layer was picked up by a lock-in amplifier. The resistance of the MG is measured from the same sample at a zero magnetic field, simultaneously.

**Table S1:** Summary of the fitting parameters of the $V_{ISHE}$ - temperature relations of both Au-based and Pd-based MGs based on a simplified two-level system model: $V_{ISHE} = V_c + V_0 \exp(-V_a/k_B T)$, where $V_c$ is the intrinsic ISHE voltage contributed by chemical and structural effects of the MGs, $V_0$ is the temperature-independent coefficient representing the interaction between spins and local excitations of MGs, and $V_a$ is the energy barrier of fast thermal relaxations of MGs.

| MG systems | $V_c$ /nV | $V_0$ /nV | $V_a$ /meV |
|---|---|---|---|
| Au-based MG | 67 | 8127 | 29.4 |
| Pd-based MG | 133 | 37371 | 37.5 |

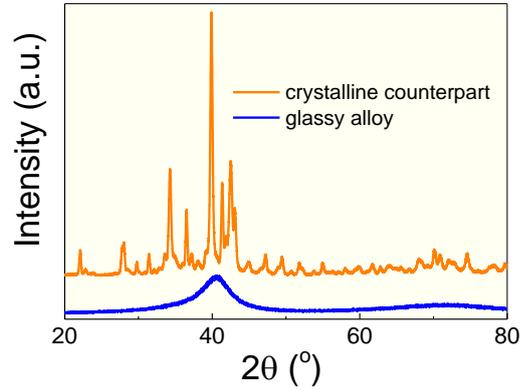

**Figure S1.** X-ray diffraction patterns of the pristine $Au_{50.5}Ag_{7.5}Cu_{26.5}Si_{15.5}$ MG and crystallized sample with the thermal experience (temperature and time) same as the 10 nm thin films heated at 400 K for spin-pumping measurements.

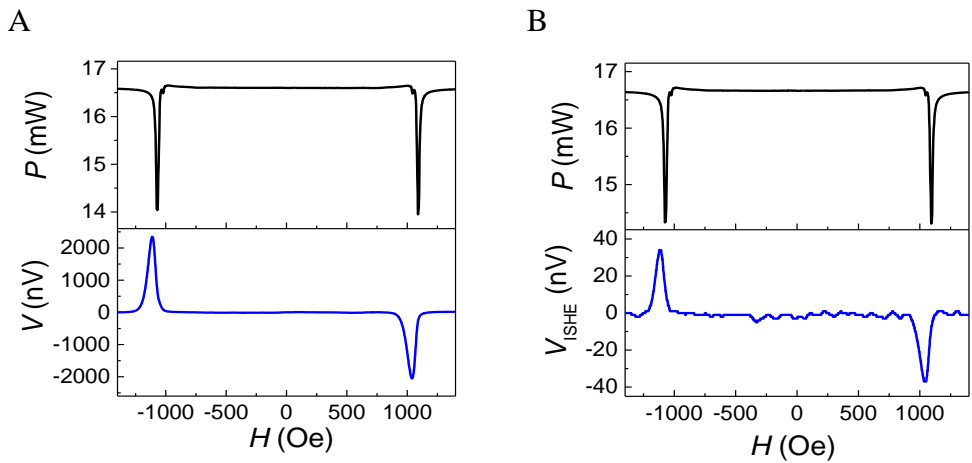

**Figure S2.** Magnetic field dependence of microwave absorption power (P) and the associated electric voltage (V) of the glassy and crystalized $Au_{50.5}Ag_{7.5}Cu_{26.5}Si_{15.5}$ alloy. (A) The MG/YIG bilayer tested at 300K during heating. (B) Crystallized Au-based MG/YIG bilayer tested at 300K during cooling. The transition from glass to crystal takes place at about 400 K during *in situ* heating/cooling spin pumping measurements.

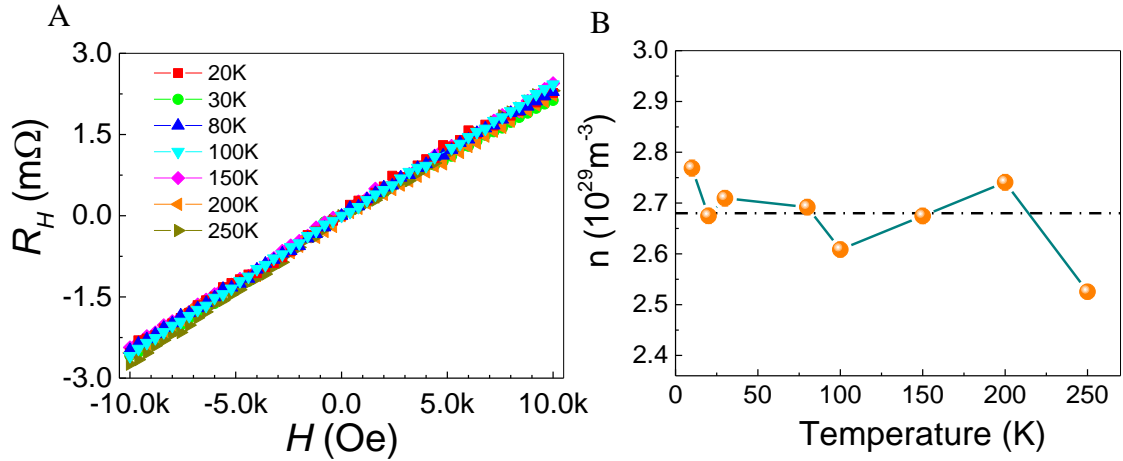

**Figure S3.** Ordinary Hall effect of the $Au_{50.5}Ag_{7.5}Cu_{26.5}Si_{15.5}$ MG. (A) The curves measured at a series of temperatures are overlapped with each other, demonstrating the Hall slope is insensitive to temperature. (B) The associated temperature dependence of electron density calculated from the Hall slope.

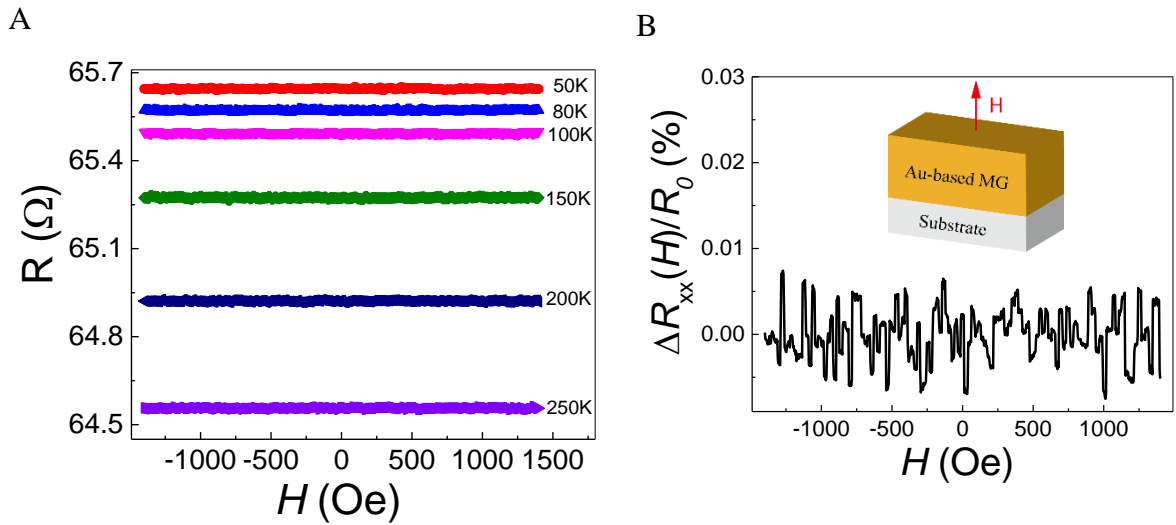

**Figure S4.** Magnetoresistance of the $Au_{50.5}Ag_{7.5}Cu_{26.5}Si_{15.5}$ MG. (A) Magnetic field dependence of resistance at a series of representative temperatures in an out-of-plane magnetic field. (B) The variation of resistance is less than 0.01% in the presence of magnetic field, as demonstrated at 200K, where $\Delta R_{xx}(H)= R(H)-R_0$. The measurement configuration is shown as the inset.

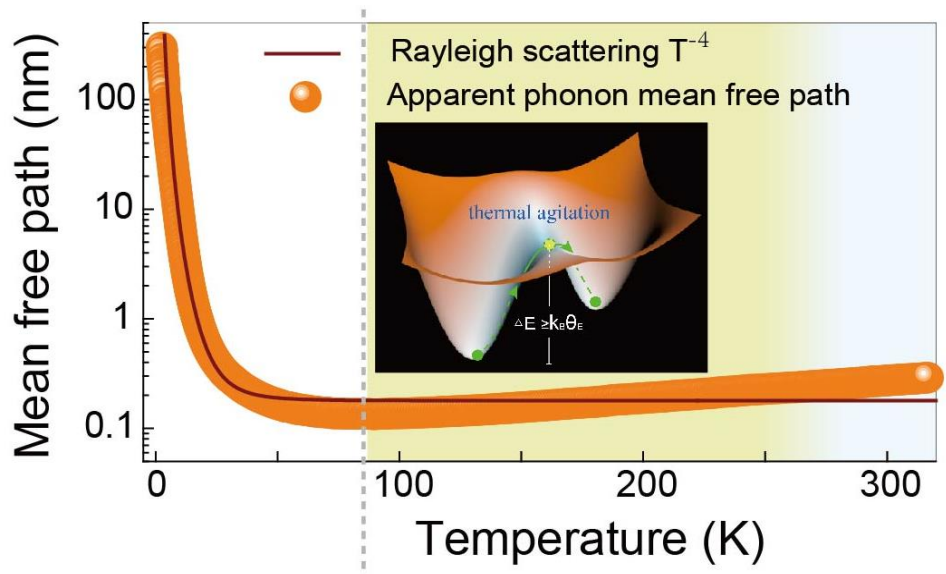

**Figure S5.** Apparent phonon mean free path in the $Au_{50.5}Ag_{7.5}Cu_{26.5}Si_{15.5}$ MG calculated from thermal conductivity and specific heat. It decreases rapidly with temperature in the Rayleigh scattering form of $T^{-4}$ at the temperatures below about 50K. The phonon mean free path becomes very short, of the order of interatomic distance, at the temperatures above 50K, demonstrating the existence of quasi-localization behavior of high frequency vibration modes in the MG. The slight increase with temperature in the high temperature range is due to the breakdown of kinetic formula of thermal conductivity and the emergence of local structural fluctuation. Inset: Schematic illustration of local structural fluctuation via thermal activation.

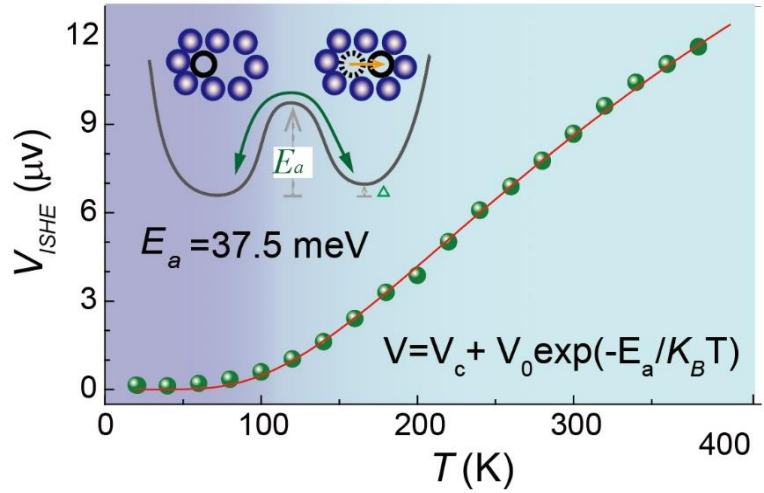

**Figure S6** The temperature dependence of $V_{ISHE}$ of Pd-based MG is fitted by a simplified two level model with the barrier energy of 37.5meV. The associated relaxation rate, calculated from $\upsilon=\upsilon_0 exp(-E_a/k_B T)$, where $\upsilon_0$ is the atomic vibration frequency in the order of $10^{12} \sim 10^{13}$Hz.

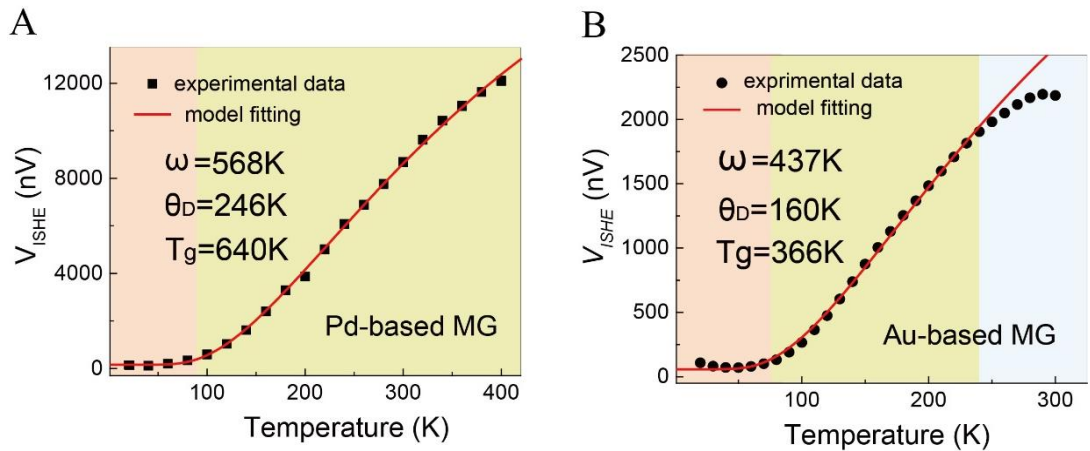

**Figure S7.** The fitting of the temperature dependence of $V_{ISHE}$ by the extended phonon skew scattering model with localized structural excitations with the frequency well above Debye temperature $\theta_D$, but close to about $T_g$ for: (A) Pd-based MG; and (B) Au-based MG. At high temperature, the deviation is associated with the onset of irreversible structural relaxation.